\documentclass[preprint]{aastex}

\usepackage{ulem}
\shorttitle{The LFs in the Hydra I Cluster}
\shortauthors{Yamanoi et al.}

\begin{document}


\title{The Galaxy Luminosity Functions down to $M \sim -10$ \\
       in the Hydra I Cluster}


\author{Hitomi Yamanoi\altaffilmark{1},
Masayuki Tanaka\altaffilmark{2}, Masaru Hamabe\altaffilmark{3}, 
Masafumi Yagi\altaffilmark{4}, Sadanori Okamura\altaffilmark{2,5}, 
Masanori Iye\altaffilmark{1,2,4},
Kazuhiro Shimasaku\altaffilmark{2,5}, Mamoru Doi\altaffilmark{6},
Yutaka Komiyama\altaffilmark{4}, Hisanori Furusawa\altaffilmark{7}}
\email{yamanoi.hitomi@nao.ac.jp}
\altaffiltext{1}{Department of Astronomical Science, School of Physical Sciences, 
The Graduate University for Advanced Studies (Sokendai), National Astronomical Observatory of Japan,
Mitaka, Tokyo 181-8588, Japan}
\altaffiltext{2}{Department of Astronomy, School of Science,
The University of Tokyo, Bunkyo-ku, Tokyo 113-0033, Japan}
\altaffiltext{3}{Department of Mathematical and Physical Sciences, School of Science, 
Japan Women's University, Bunkyo-ku, Tokyo 112-8681, Japan}
\altaffiltext{4}{Optical and Infrared Astronomy Division, National Astronomical Observatory of Japan,
Mitaka, Tokyo 181-8588, Japan}
\altaffiltext{5}{Research Center for the Early Universe, School of Science,
The University of Tokyo, Bunkyo-ku, Tokyo 113-0033, Japan}
\altaffiltext{6}{Institute of Astronomy, The University of Tokyo, Mitaka, Tokyo 181-0015, Japan}
\altaffiltext{7}{Subaru Telescope, National Astronomical Observatory of Japan, 650 N. A'Ohoku Place,
Hilo, HI 96720, USA}



\begin{abstract}

We study the galaxy population in the central region and a region about 0.6 Mpc
away from the center of the Hydra I cluster in $B$- and $R_C$-bands down
to $M\sim-10$ using the Subaru Suprime-Cam photometry.
We find that the luminosity function of the entire population has a slightly
steeper slope ($\alpha\sim-1.6$) in the range of $-20<M<-10$ than 
those reported for other clusters in slightly brighter ranges.
The slope appears to be steeper in poorer clusters.
The number of faint galaxies ($M>-14$) increases in the cluster center 
as well as bright galaxies.
The Hydra I cluster is dominated by red dwarfs and the luminosity
function shows a significant upturn at $M\sim-16$ as is seen in several
other nearby clusters, but not in the field.  
This upturn and the variation in the faint-end slope of the luminosity function 
may be caused by the cluster environment or the evolution history
of individual clusters.

\end{abstract}


\keywords{galaxies: clusters: individual: Hydra I (Abell 1060)------ galaxies: luminosity function
------ galaxies: dwarf}

\section{INTRODUCTION}

The galaxy luminosity function (LF) describes one of the fundamental statistical properties 
of galaxy populations and provides clues to the history of galaxy formation and evolution. 
The bright-end of LFs has been intensively studied for clusters of galaxies and in the field,
but the observations of the faint-end of LFs are still few.
To probe the faint-end of LFs, a large systematic spectroscopic survey is required. 
Therefore, observational studies of the faint-end of LFs are still limited to
cluster members.
Because deep photometry of clusters enables statistical evaluation 
of the faint population of cluster galaxies, 
if background/foreground contamination can be properly removed.
Dwarf galaxies account for a large share of the cluster galaxy population.
Despite their obvious importance, available observational studies of 
cluster dwarf galaxies down to $M\sim-10$ are still very limited and their properties as 
well as their role in cluster formation and evolution remain unclear.

Studies of cluster dwarf galaxies are limited to nearby clusters, i.e.,
Virgo, Fornax, Perseus, and Coma, since dwarfs are intrinsically faint.
Previous investigations of cluster LFs have shown that the number density of galaxies 
significantly increases at faint magnitudes, 
but with considerable cluster to cluster variation.
\citet{Phi98} carried out an optical survey of the Virgo cluster \citep[$z=0.0036$;][]{Ebe98} 
to estimate the number of dwarf galaxies in this cluster.
They obtained a LF with a very steep faint-end slope of $\alpha \sim-2.0$ between
$-15\lesssim M_R \lesssim-11$
(the magnitudes quoted hereafter are adjusted to the following cosmology:
$H_0=70$ km s$^{-1}$ Mpc$^{-1}$, $\Omega_m=0.3$, and $\Omega_\Lambda=0.7$).
The Fornax cluster LF \citep[$z=0.0046$;][]{Abel89} also shows a steep slope ($\alpha \sim-2.0$) 
down to $M_B\sim-12$ \citep{Kamb00}.
These results support cold dark matter models \citep{PS74, WR78, WF91, LS99},
which predict the formation of numerous low-mass dark matter halos 
(observed as low-luminosity galaxies) in the Universe.  
In contrast, \citet{Sab03} found that the Virgo LF has
a slope of $\alpha \sim -1.6$ in the magnitude range of $-15 \lesssim M_B \lesssim -11$,
which is flatter than that found by \citet{Phi98}. 
The Coma cluster \citep[$z=0.0231$;][]{SR99}, which is a rich and more distant cluster than the Virgo, 
has a faint-end slope of $\alpha \sim -1.4$ at $-19 \lesssim M_R \lesssim -11$ \citep{Ber95}.
The Perseus cluster \citep[$z=0.0179$;][]{SR99}, another rich cluster, has similar properties 
to the Coma cluster, and it shows a faint-end slope of 
$\alpha \sim -1.4$ at $M_B<-11$ \citep{Con02}.
Recently, \citet{Tre05} determined the field LF from the Sloan Digital Sky Survey (SDSS)
galaxy samples, nearby group galaxies and Local Group galaxies.
They showed that the faint-end slope of the field LF is $\alpha=-1.26$ over the range of 
$-19\lesssim M_R\lesssim -9$.

A goal of this study is to reveal the LFs down to $M\sim-10$ in the Hydra I cluster (Abell 1060)
at $z=0.0126$ \citep{SR99}, for which such a very faint galaxy population, at $M\sim-10$, 
has not been previously surveyed.
\citet{BO83} and \citet{Ric89} studied the galaxy distribution and redshifts 
of Hydra I.
They found that Hydra I is apparently isolated in redshift space and 
member galaxies of Hydra I are not heavily contaminated by background/foreground galaxies.
Therefore, this cluster is ideal for estimating LFs of member galaxies using statistical
background subtraction.
Systematic uncertainties arising from the subtraction should be smaller 
than for other cluster fields.

Many studies of LFs have revealed the presence of an abundant dwarf galaxy population 
in cluster environments.
However, most such studies have observed only central cluster regions.
The environmental dependence of properties and distribution of faint dwarfs 
remains unexplored.
If dwarfs have been tidally stripped or threshed via frequent interactions 
with bright galaxies (e.g., \citealt{Moo96, Bekk01}), the cluster core may be where 
dwarfs strongly evolve.   
In this paper, we seek to examine faint dwarf galaxies in Hydra I and to investigate 
the environmental dependence of galaxy properties by comparing LFs in the cluster center with
those in the outskirts. 

The structure of this paper is as follows.
In \S 2, we describe our observations and data reduction.
In \S 3, we derive the background galaxy number counts and color distributions of galaxies
for statistical background subtraction. 
We then present LFs of Hydra I and their environmental dependence in \S 4.
Finally, we summarize the paper in \S 5.
We adopt $H_0=70$ km s$^{-1}$ Mpc$^{-1}$, $\Omega_m=0.3$, and
$\Omega_\Lambda=0.7$.
Magnitudes are on the AB system. 
We assume a distance to the Hydra I cluster of 53.8 Mpc and a distance modulus of 33.68
throughout this paper.

%%%%%%%%%%%%%%%%%%%%%%%%%%%%%%%%%%%%%%%%%%%%%
%%%%%%%%%%%%%%%%%%%%%%%%%%%%%%%%%%%%%%%%%%%%%
%%%%%%%%%%%%%%%%%%%%%%%%%%%%%%%%%%%%%%%%%%%%%
%%%%%%%%%%%%%%%%%%%%%%%%%%%%%%%%%%%%%%%%%%%%%
%%%%%%%%%%%%%%%%%%%%%%%%%%%%%%%%%%%%%%%%%%%%%

\section{OBSERVATIONS AND DATA REDUCTION}

\begin{figure}
%\figurenum{E-1}
\epsscale{1}
\plotone{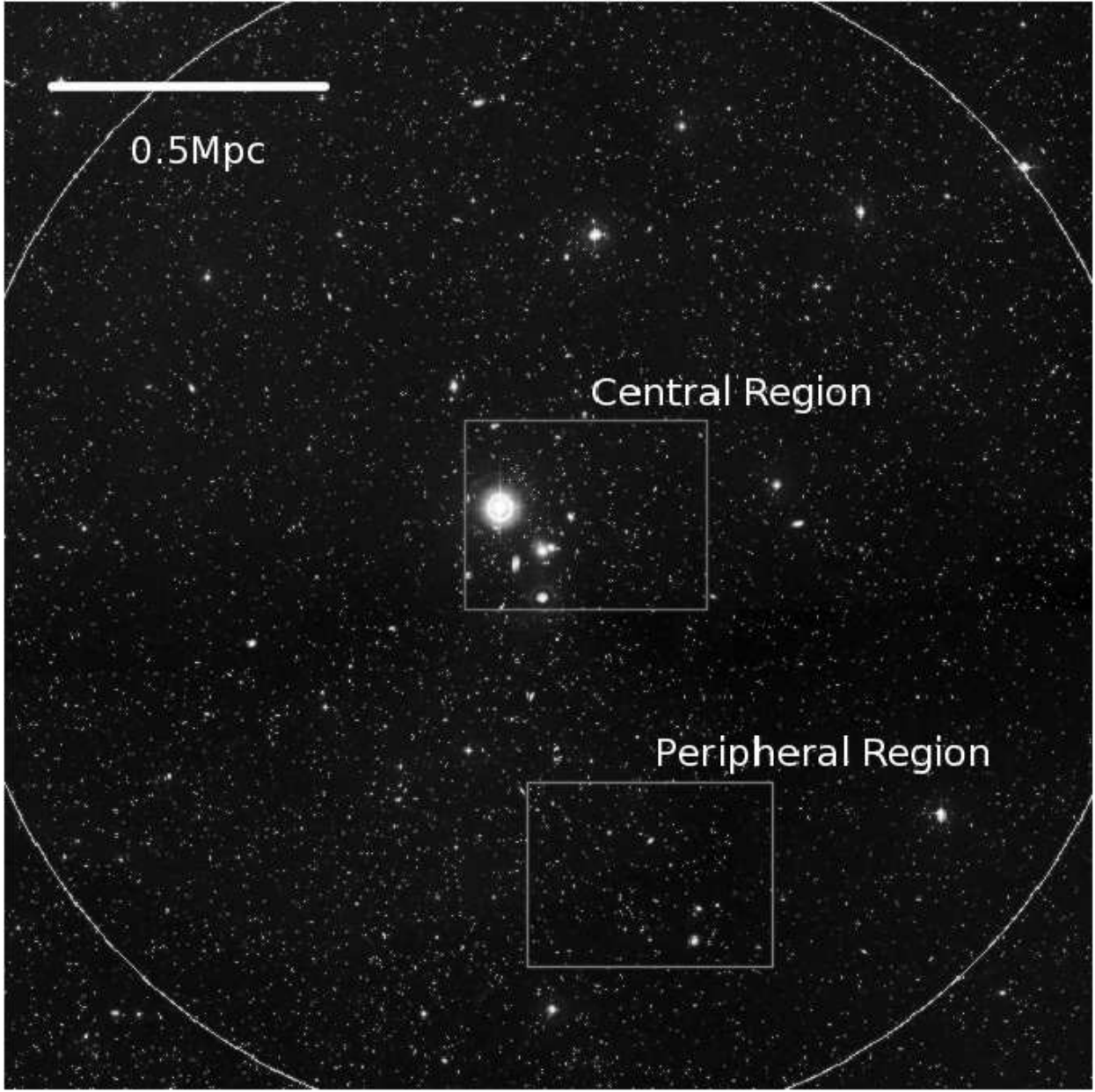}
\caption{
The Digitized Sky Survey image of the Hydra I cluster. North is in the upward direction.
The rectangles show the central and peripheral regions.
The circle indicates the virial radius ($r_{200}$) of the cluster estimated 
from its velocity dispersion \citep{SR91}.
The 0.5 Mpc scale is shown at the top-left.
\label{fig:hydra_map}
}
\end{figure}

Observations were made in March 2001 and November 2002.
To probe very faint dwarf galaxies 
in the Hydra I cluster, we use the Subaru Prime Focus Camera
\citep[Suprime-Cam;][]{Miya02} on the Subaru telescope \citep{Iye04} at Mauna Kea, 
which covers a $34' \times 27'$ field of view.
We observed the central and peripheral regions of 
the Hydra I cluster (see Figure \ref{fig:hydra_map}) in $B$ and $R_C$ to investigate LFs
in the different regions of the cluster. 
The peripheral region is located at about 0.6 Mpc projected distance to the south of the central
region.
For each region, we obtained short- and long-exposure images to cover a wide luminosity
range of sample galaxies.
The observational data are summarized in Table \ref{tbl-1}.

The images were reduced using the dedicated reduction software for Suprime-Cam
\citep{YK02a, Ouc04} and IRAF.
Photometric standard fields, SA101 and SA107 \citep{Lan92}, were observed to establish
the photometric zero point during the run.
We obtain accurate zero points for deep images, which were taken in March 2001
using the standard stars.
We determine zero points for shallow images taken in November 2002
using relative photometry.
To convert from the Vega system to the AB magnitude system, we adopt the following transformations: 
$B{\rm(AB)}=B{\rm(Vega)}-0.140$ and $R_C{\rm(AB)}=R_C{\rm(Vega)}+0.169$ 
\citep[see][]{Fuku96}.
The seeing size was about 1.1 arcsec throughout the observations.
We use SExtractor \citep{BA96} to detect objects and 
adopt \verb|MAG_AUTO| for total magnitudes and 2 arcsec aperture magnitudes for colors. 
Regions around bright stars and giant galaxies, 
where the detection of faint objects fails, are masked out.
The effective areas of the central and peripheral regions in Hydra I are
$422$ arcmin$^2$ and $498$ arcmin$^2$, respectively. 

Star-galaxy separation is performed on the basis of FWHM versus total magnitude diagrams.
Note that we may miss extremely compact objects such as  ultracompact dwarfs \citep{Drink03};
however, they comprise a very small fraction of member galaxies and we can ignore
their contribution to LFs \citep{AC02, Drink03}.
The Galactic extinction is corrected using the extinction map of \citet{Sgel98}.
The 5$\sigma$ limiting magnitudes measured within 2 arcsec apertures
are summarized in Table \ref{tbl-1}.
Because of the relatively poor seeing size of our data, star/galaxy separation is difficult
at magnitudes fainter than $m_R\sim24.0$.
We calculate LFs in a wide range of magnitudes $12.7<m_R\leq23.7$.
The final composite $B$- and $R_C$-band images are sufficiently deep (we reach $M_R=-10$) 
to study low-luminosity dwarf galaxies.
The $R_C$-band is expressed as the $R$-band for simplicity in what follows.

%%%%%%%%%%%%%%%%%%%%%%%%%%%%%%%%%%%%%%%%%%%%%%%%%%%%%%
%%%%%%%%%%%%%%%%%%%%%%%%%%%%%%%%%%%%%%%%%%%%%%%%%%%%%%
%%%%%%%%%%%%%%%%%%%%%%%%%%%%%%%%%%%%%%%%%%%%%%%%%%%%%%
%%%%%%%%%%%%%%%%%%%%%%%%%%%%%%%%%%%%%%%%%%%%%%%%%%%%%%

\section{CONTAMINATION SUBTRACTION}
\label{sec:background}

To obtain the intrinsic LFs of cluster galaxies, we must correct for the number distribution
of contaminant galaxies, which are mainly located behind the cluster.
We use three different methods for subtracting the contamination galaxies
depending on magnitude.
For the bright magnitude range, we use spectroscopic data from the literature \citep{Ric82, CZ03};
for the middle magnitude range, we use data from the Sloan Digital Sky Survey
\citep[SDSS;][]{Yor00} to perform statistical subtraction;
and for the faint magnitude range, we use data from the Subaru/XMM-Newton Deep Survey Field 
(SXDS; Sekiguchi et al. in prep.)

%%%%%%%%%%%%%%%%%%%%%%%%%%%%%%%%%%%%%%%%%%%%%%%%%%%%%%
\subsection{Bright Magnitude Range: Use of Spectroscopic Data}

First, we use spectroscopic data from the literature \citep{Ric82, CZ03},
which are taken from the NASA/IPAC Extragalactic Database (NED).
Although the spectroscopic identification of member galaxies is an effective way of eliminating 
contamination, spectral redshifts are not available for some galaxies.
We estimate the number of bright Hydra I members as follows.
For a given magnitude bin, the number of member galaxies is given by:
\begin{equation}
\label{eq:estmate_member}
N_{mb}=N_{spec.mb}+\frac{N_{spec.mb}}{N_{spec.mb}+N_{spec.n\mathchar`- mb}}N_{n\mathchar`- spec}, 
\end{equation}
where $N_{spec.mb}$ is the number of member galaxies identified with spectroscopic redshifts,
$N_{spec.n\mathchar`- mb}$ is the number of nonmember galaxies with spectroscopic redshifts,
and $N_{n\mathchar`- spec}$ is the number of galaxies with no redshifts.
The second term gives the expected number of member galaxies in $N_{n\mathchar`- spec}$.
We apply this estimation for a $12.7<m_R<17.7$ magnitude range in the central region of 
Hydra I to identify the member galaxies.
In the peripheral region of Hydra I, we adopt a magnitude range of $12.7<m_R<15.7$,
somewhat shallower than that of the central region due to the poor spectroscopic sampling.

%%%%%%%%%%%%%%%%%%%%%%%%%%%%%%%%%%%%%%%%%%%%%%%%%%%%%%
\subsection{Middle Magnitude Range: Use of SDSS Data}
 
Second, we use data from the SDSS to estimate
the number of contaminant galaxies at a given magnitude for statistical subtraction.
The SDSS covers a quarter of the whole sky and such large survey data are suited
for statistical subtraction of contamination since the effects of large-scale structures
are minimized.
We extract photometric data from the public data release four (DR4; \citealt{AM06}).
We then transform the SDSS photometry into $B$ and $R$ magnitudes using the recipe
provided on the SDSS Web site (Lupton's equation).
The magnitude distribution of galaxies is normalized to the effective surface area of the central 
and the peripheral regions.
We evaluate the number of contaminant galaxies in each field at a given magnitude bin
and statistically subtract them to obtain a LF of cluster members.
We apply this statistical contamination subtraction for magnitude ranges of
$17.7<M_R<19.7$ and  $15.7<M_R<19.7$ for the central and peripheral regions, respectively.

To evaluate the effects of cosmic variance,  
we randomly place an aperture with areas equivalent to those 
for the Hydra central or peripheral region to extract galaxies in the SDSS survey region. 
We perform the galaxy number count 100,000 times and evaluate the cosmic variance.
The cosmic variance is found to be comparable to the Poisson error within $\pm10\%$. 
We therefore have confirmed that our discussions presented below are unaffected by 
cosmic variance.

%%%%%%%%%%%%%%%%%%%%%%%%%%%%%%%%%%%%%%%%%%%%%%%%%%%%%%
\subsection{Faint Magnitude Range: Use of SXDS Data}

Third, we use public data from the SXDS archive, 
which is a wide and deep survey of an approximately one square degree region of the sky.
The SXDS is a blank field survey (i.e., no nearby clusters) and
therefore it is suited for estimating the population of faint background galaxies
for subtraction.
Note that the imaging data of the SXDS are taken with the same filters as our data 
and no additional error is introduced from band transformation.
The 5$\sigma$ limiting magnitudes measured within 2 arcsec apertures
are $m_B=27.5$ and $m_R=26.8$ respectively.
The seeing size was 0.7 - 0.8 arcsec. 
The Galactic extinction is estimated based on \citet{Sgel98}.
We use the results from this SXDS field for the statistical subtraction of 
the galaxy background in the fainter magnitude range of $19.7<m_R<23.7$ for both the central 
and the peripheral regions.

%%%%%%%%%%%%%%%%%%%%%%%%%%%%%%%%%%%%%%%%%%%%%%%%%%%%%%
\subsection{The Galaxy Number Counts}

\begin{figure}
%\figurenum{E-2}
\epsscale{1}
\includegraphics[angle=270,scale=.67]{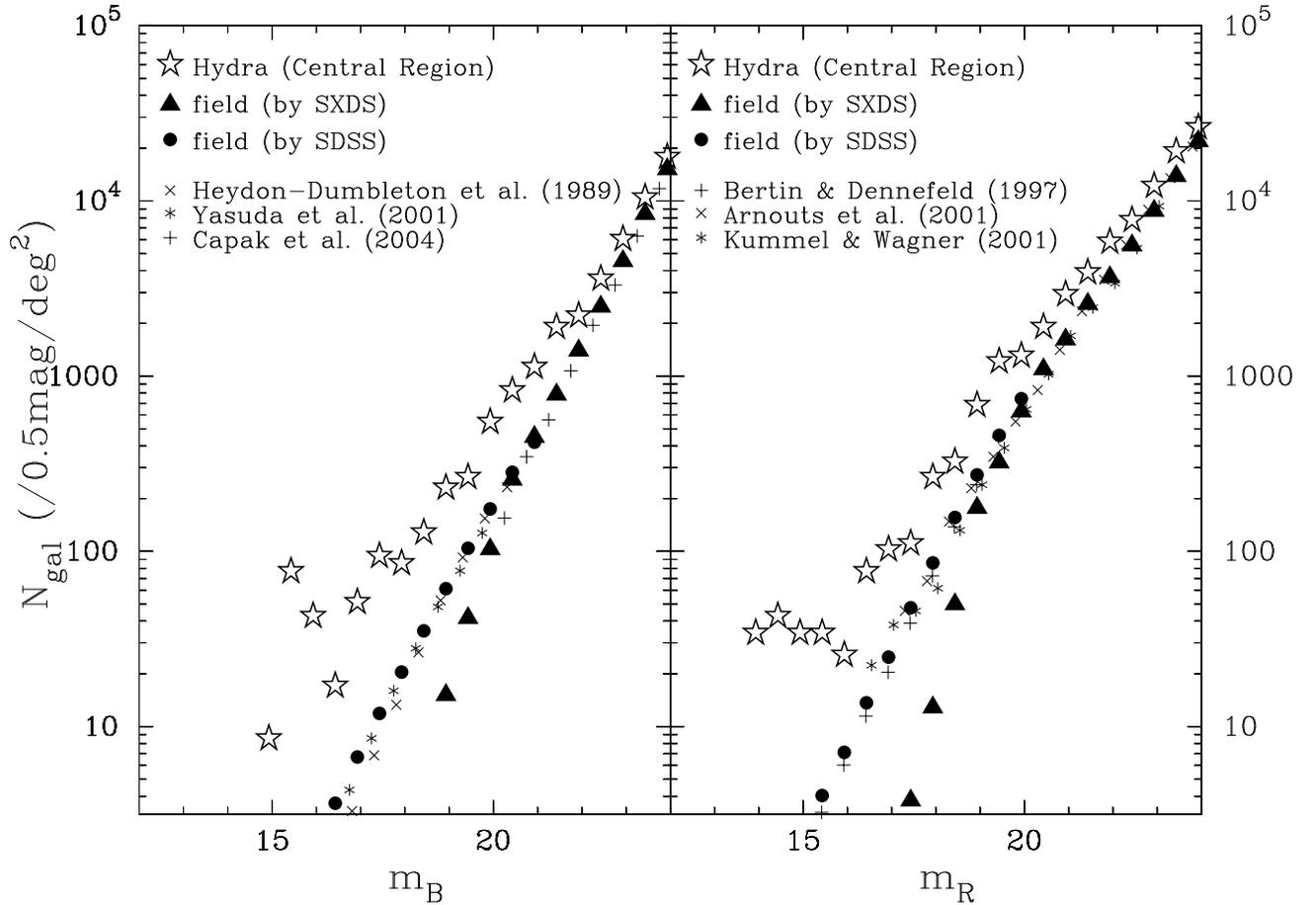}
\caption{
Number counts of galaxies as a function of magnitude in the $B$- (left) and $R$-bands (right).
The stars indicate the galaxy counts of the central region in the Hydra I cluster.
The triangles and circles show the SXDS and SDSS galaxy counts, respectively. 
A lack of triangles at bright magnitudes ($m_B<20.0$ and $m_R<19.5$) is due to image
saturation.
\label{fig:logn_vs_m}
}
\end{figure}

We check whether the background galaxy number counts from the SDSS and SXDS data are consistent
in the overlapping magnitude region.
Figure \ref{fig:logn_vs_m} shows the galaxy number counts in 0.5 mag bins 
in the $B$- and $R$-bands in other fields from the literature as well as 
the Hydra central region.
At $m_B\sim21$ and $m_R\sim20$, we see the consistency of the field galaxy number counts
from the SDSS and SXDS data.
Compared to other field galaxy samples from previous studies 
(e.g., \citealt{HD89, BD97, Arn01, Yasu01, KW01}),
the galaxy number counts that we calculate using the SDSS and SXDS data are consistent at
$16.0<m_B<24.0$ and $15.0<m_R<24.0$.
At bright magnitudes ($m_B<20.0$ and $m_R<19.5$), the SXDS data points (triangles)
fall below the SDSS data and the other works.
This is caused by the saturation of bright galaxies in the SXDS field.
At faint magnitudes ($20.5<m_B<24.0$ and $20.0<m_R<24.0$), the SXDS points are consistent with
\citet{Capa04}, \citet{Arn01} and \citet{KW01}.

To discuss the environmental dependence of galaxy properties,
we separate the Hydra member galaxies into red and blue galaxies.
Our prescription to divide red and blue galaxies is described below.
We measure galaxy colors within 2 arcsec apertures for the Hydra, SDSS, and SXDS samples.
For SDSS, we derive 2 arcsec aperture colors from radial profiles
of galaxies (PhotoProfile in the Catalog Archive Server).
The SDSS observations are made under various conditions and the aperture photometry
may not be suitable for measuring galaxy colors in SDSS galaxies.
To examine the effect of aperture sizes, we adopt a color measured for the entire galaxy
('model' colors) and a color in 1.3 arcsec apertures and compare how our results change.
In fact, the results are essentially unchanged.
We therefore adopt 2 arcsec aperture colors for all the samples.
 
\begin{figure}
%\figurenum{E-3}
\epsscale{1}
\includegraphics[angle=270,scale=.90]{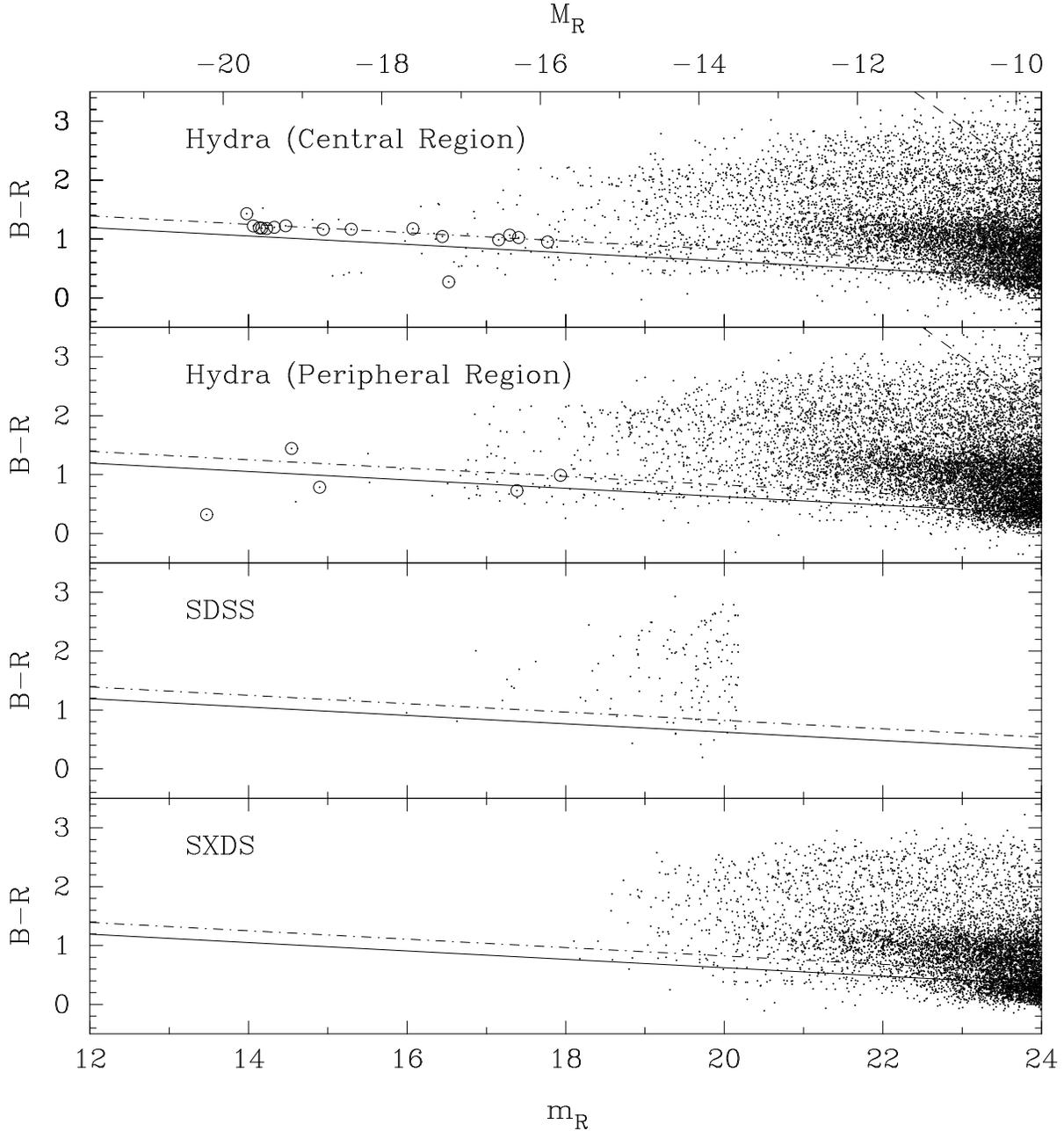}
\caption{
The CMDs ($B-R$ vs. $m_R$ ) in the Hydra I cluster.
Galaxies in the central region, the peripheral region, the SDSS and SXDS are 
plotted from top to bottom.
The open circles are the spectroscopic member galaxies.
The dot-dashed lines represent the CMR. 
We separate red and blue galaxies based on this line shifted by $\Delta(B-R)=-0.2$ as 
indicated by the solid line.
The slanted dashed lines show the 5$\sigma$ limiting color. 
Note that in the SDSS and SXDS plots, the numbers of galaxies are normalized to 
the effective surface area of the central region, and galaxies are randomly extracted 
from photometric catalogs and plotted.
\label{fig:cmd}
}
\end{figure}

We present color-magnitude diagrams (CMDs) of Hydra I and the control fields 
(SDSS and SXDS) in Figure \ref{fig:cmd}.
The open circles in Figure \ref{fig:cmd} are the spectroscopic member galaxies.
The lines indicate the least-squares fit to the color-magnitude relation (CMR) of 
early-type (E/S0) member galaxies in the center of Hydra I.
The fitted relation is
\begin{equation}
\label{eq:CMR}
(B-R)=(-0.071\pm0.012)m_R+(2.25\pm0.19).
\end{equation} 
In addition to deriving the LF of all member galaxies (total LF), we derive LFs of
red and blue galaxies.
We separate the member galaxies of Hydra I into red and blue galaxies based on their 
color-magnitude relation shifted by $\Delta(B-R)=-0.2$.
(Note that our results remain unchanged if we define the red/blue separation as 
$\Delta(B-R)=-0.1$.)
This red/blue separation is based on the assumption that the color-magnitude
relation is linear down to very faint magnitudes,
and this relation is observed to be linear at least down to $M_R=-14$ \citep{And06}.
%It seems that the luminosity-metallicity relation holds even at $M_B\sim-11$ \citep{Lee06}.
The results of \citet{Adam06} suggest that low-luminosity galaxies down to 
$M_R\sim-11$ in the Coma cluster ($z=0.0231$) still follow the CMR.
Therefore, the assumption appears reasonable.
We thus have confirmed that the number counts of red and blue galaxies from the SDSS and SXDS data
are consistent in the overlapping magnitude regions.

%%%%%%%%%%%%%%%%%%%%%%%%%%%%%%%%%%%%%%%%%%%%%
%%%%%%%%%%%%%%%%%%%%%%%%%%%%%%%%%%%%%%%%%%%%%
%%%%%%%%%%%%%%%%%%%%%%%%%%%%%%%%%%%%%%%%%%%%%
%%%%%%%%%%%%%%%%%%%%%%%%%%%%%%%%%%%%%%%%%%%%%
%%%%%%%%%%%%%%%%%%%%%%%%%%%%%%%%%%%%%%%%%%%%%

\section{RESULTS AND DISCUSSION}
\label{sec:results}

A LF describes a magnitude distribution of galaxies and
provides a convenient index for comparing of galaxy populations in different environments.
\citet{Sch76} proposed a practical analytic expression for a galaxy LF.
The Schechter function has three parameters: 
the characteristic absolute magnitude $M^{*}$, the faint-end slope $\alpha$, 
and the normalization number density $\phi^{*}$.
The LF of nearby clusters (including the Hydra I cluster) is found to have a
characteristic magnitude of $M^{*}_R\sim-20.5$ \citep{YK02b, CZ03}.
We derive LFs over a magnitude range of $-20<M_R<-10$.
We focus on the faint-end of LFs and quantify LFs with only one parameter, $\alpha$.   
We examine (1) total (red $+$ blue) LFs and (2) red/blue LFs 
in the central and the peripheral regions.

%
%
%%%%%%%%%%%%%%%%%%%%%%%%%%%%%%%%%%%%%%%%%%%%%%%%%%%%%%
\subsection{Total LFs in the Hydra I Cluster}

\begin{figure}
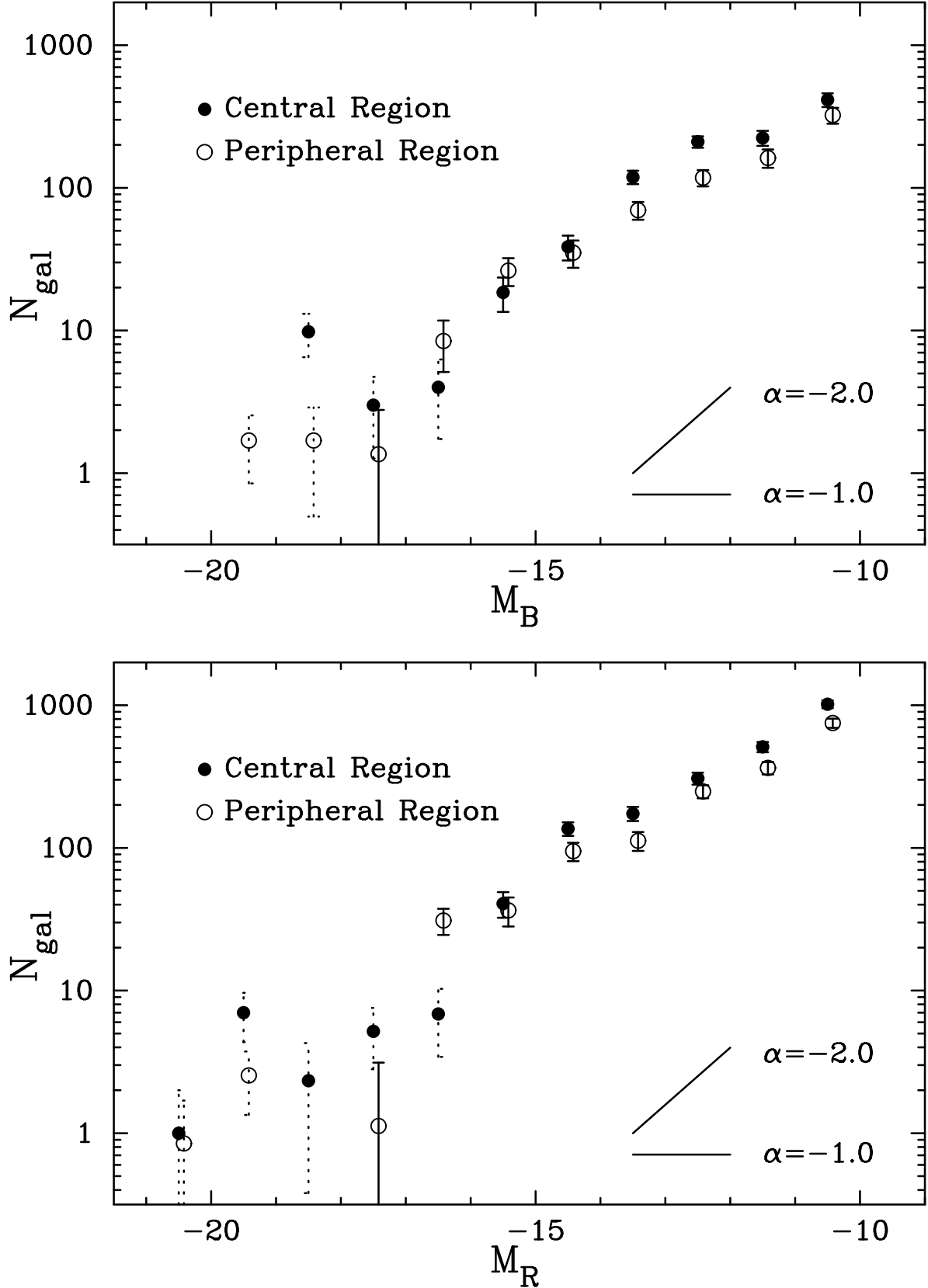

%\figurenum{E-4}
\epsscale{1}
\begin{center}
\includegraphics[angle=270,scale=.67]{f4_a.ps}\\
\vspace{0.5cm}
\includegraphics[angle=270,scale=.67]{f4_b.ps}
\caption{
Total LFs in the central region (filled circles) and the peripheral region (open circles) 
in the Hydra I cluster.
The top and bottom panels show the $B$- and $R$-band LFs, respectively.
The error bars are based on Poisson statistics.
Note that the circles with the dotted error bars are derived using the spectroscopic samples 
(see \S 3 for details). 
\label{fig:total_LF}
}
\end{center}
\end{figure}

We obtain the LFs of the Hydra I cluster down to $M\sim-10$
for the $B$- and $R$-bands as shown in Figure \ref{fig:total_LF}.
Note that LFs of the peripheral region are normalized to the effective surface area of 
the central region.

We fit the LFs with a power-law model.  
The best-fit models have $\alpha=-1.59\pm0.03$ ($-19.0<M_B<-10.0$)
and $\alpha=-1.64\pm0.02$ ($-21.0<M_R<-10.0$) in the central region in the entire magnitude range.
In the peripheral region, the best-fit slopes are $\alpha=-1.61\pm0.03$ ($-20.0<M_B<-10.0$) and 
$\alpha=-1.62\pm0.03$ ($-21.0<M_R<-10.0$), respectively.
The slope, $\alpha\sim-1.6$, is steeper than the faint-end slope of the composite LF 
in nearby clusters (including Hydra I) reported by \citet{YK02b}, 
$\alpha=-1.31$ at $-22<M_R<-15$, or \citet{CZ03}, $\alpha = -1.21$ at $-22<M_R<-16$.
It should, however, be noted that we reach $\sim5$ mag deeper than the previous studies. 
As shown in Figure \ref{fig:total_LF}, the LFs are steeper ($\alpha\sim-1.6$) at $M_R\gtrsim-16.5$, 
while the LFs are flatter ($\alpha\sim-1.0$) at $M_R\lesssim-16.5$.
We conclude that the faint-end slope $\alpha$ in the LFs of Hydra is 
rather steep ($\alpha\sim-1.6$) at a faint magnitude of $M\gtrsim-16$ 
in the $B$- and $R$-bands.

\citet{Pop05} found an upturn of the dwarfs and steepening of the LFs at the faint-end 
in the five SDSS photometric bands for 114 clusters in the redshift range 0.002-0.45.
\citet{Tre05} reported on a composite LF in nearby clusters over a range of $-25<M_R<-9$ and
also found the upturn of the LF at $M_R\sim-16$. 
In Figure \ref{fig:total_LF}, the upturn due to dwarfs is also seen at $M\sim-16$, which is 
in broad agreement with previous studies.
Note that the upturn of the faint galaxies in Hydra I appears at the magnitude threshold 
at which we change the contamination subtraction scheme.
As a consistency check, we carry out the statistical background subtraction for galaxies at 
$-20.0<M_B<-13.0$ and $-21.0<M_R<-14.0$ using the SDSS data set only.
We again find that our LFs show the upturn at $\sim-16$ mag in the $B$- and $R$-bands.
Therefore, the upturn is not caused by statistical error in the background subtraction,
but is a real trend.

\begin{figure}
%\figurenum{E-5}
\epsscale{1}
\plotone{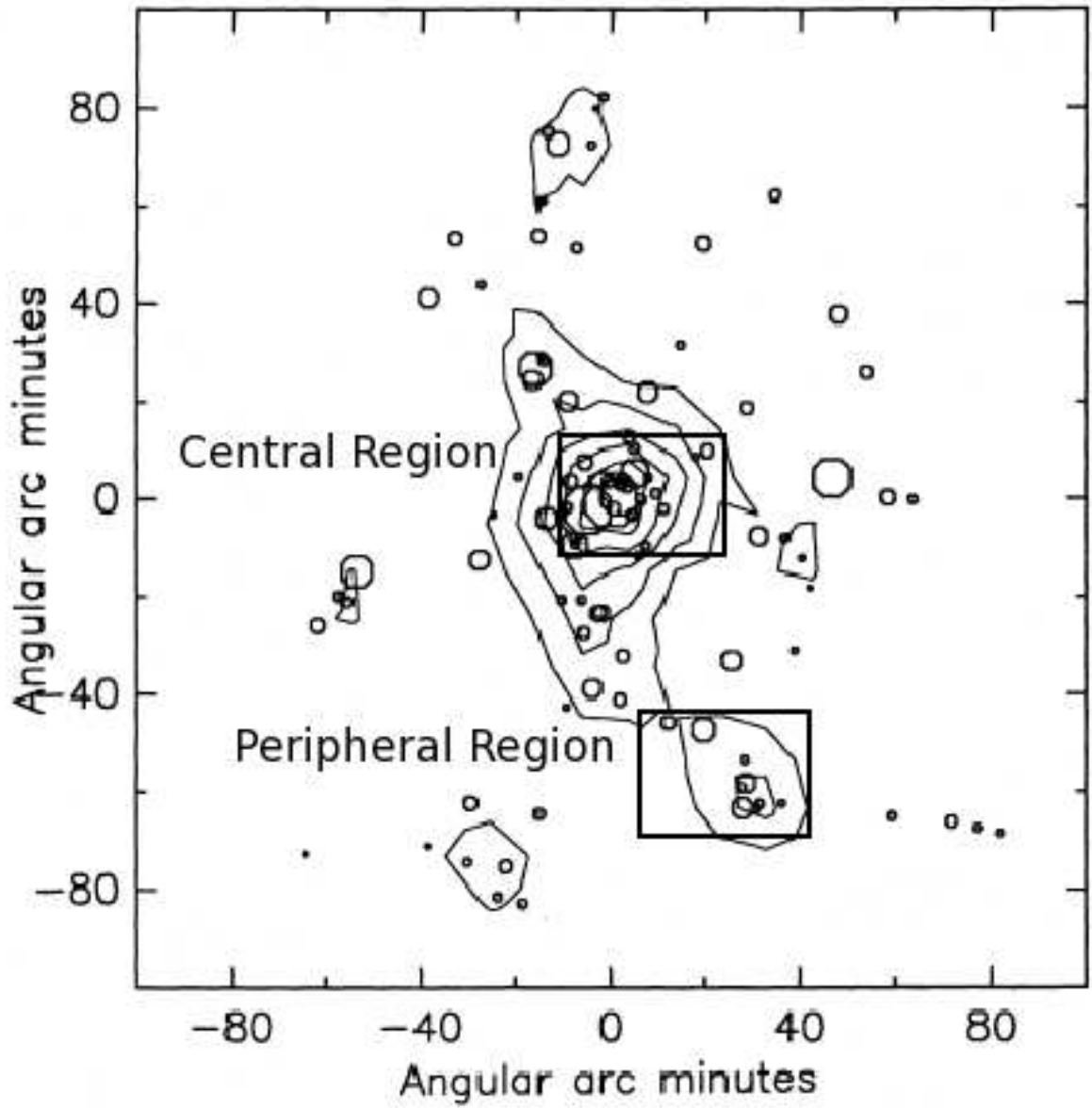}
\caption{
Contour map of the number density of galaxies in the Hydra I cluster by \citet{FM88}.
The size of each circle is proportional to the luminosity of galaxies from \citet{Ric87}.
The rectangles show the central and peripheral regions, respectively.
\label{fig:contour}
}
\end{figure}

We find no significant difference in the slopes of LFs in the different environments 
of the central and peripheral regions.
The number of bright galaxies in the LFs of the central and peripheral regions 
at $M_B<-16$ and $M_R<-15$ are not different within the errors. 
However, the numbers of galaxies at the faint-end ($M_B>-16.0$, $M_R>-15.0$) differ  slightly
between the central and peripheral regions.
The number of faint galaxies at $M_B>-16.0$ or $M_R>-15.0$ in the central region is larger 
by $37\%$ compared to that in the peripheral region.
Giant galaxies are known to be more common in the cluster center \citep{DeP03}.
We find that the numbers of dwarf galaxies also increase in the cluster center.
We note that the peripheral region accidentally includes the local peak of 
galaxy distribution as shown in Figure \ref{fig:contour} \citep{FM88}.
Accordingly, the number of galaxies in the peripheral region must be larger than 
the average for the regions at the same radial distance from the center.
The value of $37\%$ should therefore be considered as a lower limit.

We compare the LFs of the Hydra central region with those in nearby clusters from the literature. 
The total LF of the Coma cluster has a faint-end slope of $\alpha\sim-1.41\pm0.05$
at $M_R<-12$ \citep{Sec97}.
\citet{Con02} found a faint-end slope of $\alpha=-1.44\pm0.04$ in the Perseus cluster
in the magnitude range $M_B<-11$.
The LF in the Fornax cluster indicates a rather steep slope ($\alpha \sim-2.0$) 
down to $M_B\sim-12$ \citep{Kamb00}.
\citet{Sab03} reported that the faint-end slope of the LF in the Virgo cluster
is $\alpha=-1.6$ to $-1.7$ ($-15\lesssim M_B\lesssim -11$).
Note that the slope of the LF in the Hydra cluster remains constant ($\alpha\sim-1.6$)
at $-16<M<-10$ (see Table \ref{tbl-2}). 
The LF slopes may vary from cluster to cluster at such faint magnitudes.

We summarize the faint-end slopes for nearby clusters in Table \ref{tbl-2}.
To examine the correlation between cluster richness and faint-end slope ($\alpha$),
we also show the X-ray luminosities ($L_X$).
It appears that $\alpha$ increases with $L_X$.
The LFs derived by previous studies come from various regions of the cluster.
Some probed only the cluster cores, while others probed the cores as well as outskirts.  
The observed variation might be caused by the variation in the area coverage.  
However, the difference in $\alpha$ between the central and peripheral region in Hydra I 
is very small ($\Delta\alpha\sim0.02$).
It is therefore unlikely that the variation in area coverage causes
the large variance in $\alpha$, 
if $\alpha$ variation between the central region and the
outskirts of other clusters in general is as small as that observed in Hydra I.  
Although it is premature to draw any physical interpretation from the $\alpha$ variation
among different clusters,
it could be related to the formation histories of clusters as discussed in the next subsection. 

%
%   
%%%%%%%%%%%%%%%%%%%%%%%%%%%%%%%%%%%%%%%%%%%%%%%%%%%%%%
\subsection{LFs of Red/Blue Galaxies in the Hydra I Cluster}

We study the LFs of red and blue galaxies in the two different regions to examine the
environmental dependence for galaxy colors.
We divide galaxies into red and blue galaxies on the basis of their $B-R$ color 
and derive LFs in the same way as the total LFs.

\begin{figure}
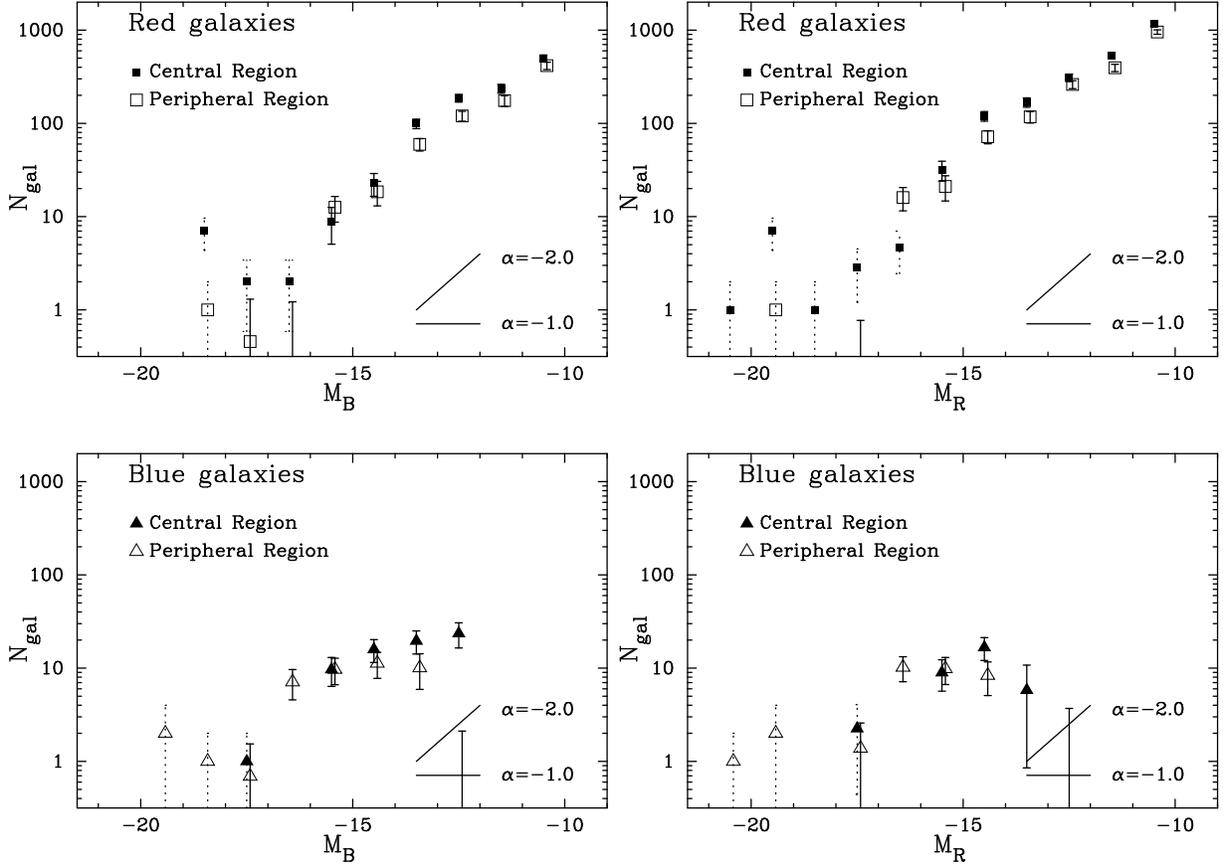

%\figurenum{E-6}
\epsscale{1}
\begin{center}
\includegraphics[angle=270,scale=.34]{f6_a.ps}
\includegraphics[angle=270,scale=.34]{f6_b.ps}\\
\vspace{0.5cm}
\includegraphics[angle=270,scale=.34]{f6_c.ps}
\includegraphics[angle=270,scale=.34]{f6_d.ps}
\caption{
Red and blue LFs in the central region (filled symbols) and the peripheral region (open symbols)
in the Hydra I cluster.
The top and bottom panels denote the red and blue LFs, respectively.
The left and right panels show the $B$- and $R$-band LFs, respectively.
The error bars are determined in the same way as for Fig. \ref{fig:total_LF}.
Note that blue LFs suffer from the uncertainty in the background subtraction
at the very faint magnitude range ($M_B>-13$, $M_R>-14$).
\label{fig:rb_LF}
}
\end{center}
\end{figure}

The red and the blue LFs of Hydra I are shown in Figure \ref{fig:rb_LF}.
We find that the red LFs are similar to the total (red$+$blue) LFs (see Fig. 
\ref{fig:total_LF}) in both $B$- and $R$-band.
The best-fit line of the red LF is $\alpha\sim-1.7$ in the central and the peripheral region 
at overall magnitudes.
The number of red faint galaxies ($M_R>-16$)in the central region is 
about $30\%$ larger than that in the peripheral region.
\citet{FS88} also reported that an excess of red dwarfs is seen down to $M_{B_T}\sim-12$
in the cluster region for Fornax and Virgo compared to the outskirts.
The blue LFs in the central and peripheral regions are similar, although
they suffer from uncertainty in the background subtraction
at very faint magnitudes ($M_B>-13$, $M_R>-14$).

The number of red dwarf galaxies in both the central and the peripheral regions is large 
compared to blue dwarf galaxies.
The ratio of red to blue galaxies is $2.9$ at $-16<M_B<-13$ or $5.9$ at $-16<M_R<-14$
in the central region, while it is $2.9$ at $-16<M_B<-13$ or $5.1$ at $-16<M_R<-14$ 
in the peripheral region.
The number of red dwarf galaxies is more than $\sim3$ times larger than the number of blue ones
and are the dominant population in clusters even at $M\sim-10$.  

In Figure \ref{fig:rb_LF}, we cannot confirm a clear excess of blue dwarf galaxies in 
the central region compared to the peripheral region, unlike for red dwarfs.
The number fraction of blue galaxies in the central region is slightly larger than 
that of the peripheral region,
but the difference is modest, less than $10\%$ at $-21<M<-13$. 
Thus, it appears that the population of blue dwarf galaxies does not strongly depend on
environment, unlike the red population in the Hydra I cluster.
\citet{DeP03} examined the environmental dependence of
LFs of giant galaxies ($M_{b_J}<-16$) based on the 2dF data.
It is interesting that they also found that environmental dependence of LFs is different 
for red and blue galaxies.
Therefore, properties of galaxies may depend on the environment over 
a wide magnitude range of $M\lesssim-10$.

An excess of red galaxies over blue ones in the center of the cluster is probably 
associated with gravitational encounters with brighter galaxies and tidal heating by  
'galaxy harassment' \citep{Moo96, Moo98}. 
This process is effective for disk-dominated galaxies.
Harassment may destroy the disk structure and transform these galaxies into faint dwarf galaxies.
Star-formation activity should be significantly weakened during interactions.
We find that the red to blue galaxy ratio in the peripheral region is about $30\%$ - $40\%$ 
smaller than that in the central region.
High-density environments such as the centers of clusters may be more effective in 
transforming blue galaxies into red ones.

\citet{Hil03} found an upturn in the LF of red dwarf galaxies 
in the Fornax cluster at $M_V\sim-14$.
The upturn of LFs from $-14$ to $-16$ mag is also seen in the Coma and 
the Perseus cluster (see \citealt{Ber95} and \citealt{Con02}, respectively).
No such upturn is observed in the field \citep{Jone06, Bla05, Tre05}.
This suggests that the upturn is unique to cluster environments.
We find that red galaxies dominate over blue galaxies in the Hydra I cluster down to $M=-10$.
Therefore, the upturn is likely due to the generation of faint red galaxies, which are presumably
related to cluster environments.
Tidal interactions such as harassment should occur frequently in clusters,
and many galaxies are probably transformed into red galaxies, which
possibly form the observed upturn.  
However, to firmly identify the cause of the upturn, 
it is necessary to compare red dwarfs around $M\sim-14$ in clusters with those in the field.  
We need more detailed information about such dwarfs to verify the scenario we suggest here.  
We plan to carry out a spectroscopic survey of such dwarfs.

The number of dwarf galaxies should also be closely related to
the formation histories of clusters.
Cluster-cluster mergers \citep[e.g.][]{MO03,Owen05} may have a strong impact on the production of dwarf galaxies
due to strong dynamical interactions.
If a cluster is formed relatively quietly through a gradual accretion of field galaxies,
we may expect the number of dwarf galaxies to be small.
However, if a cluster is built up with vigorous cluster-cluster mergers,
a large number of dwarf galaxies may be seen.
The variation of the very faint-end could reflect a variation in cluster formation histories.

%%%%%%%%%%%%%%%%%%%%%%%%%%%%%%%%%%%%%%%%%%%%%
%%%%%%%%%%%%%%%%%%%%%%%%%%%%%%%%%%%%%%%%%%%%%
%%%%%%%%%%%%%%%%%%%%%%%%%%%%%%%%%%%%%%%%%%%%%
%%%%%%%%%%%%%%%%%%%%%%%%%%%%%%%%%%%%%%%%%%%%%
%%%%%%%%%%%%%%%%%%%%%%%%%%%%%%%%%%%%%%%%%%%%%

\section{SUMMARY}
\label{sec:summary}

In this paper, we investigate the $B$- and $R_C$-band LFs in the Hydra I cluster 
over a magnitude range from $\sim-20$ down to $\sim-10$ in its central and 
peripheral regions to examine the environmental dependence of galaxy properties.

Our primary findings are as follows.

\begin{itemize}
\item We find that the faint-end slopes of LFs in the Hydra I cluster are 
$\alpha \sim -1.6$ at the magnitude range of $-20<M<-10$ in both $B$- and $R_C$- bands.
The faint-end slopes of the LF of other nearby clusters, e.g., Virgo, Fornax, Perseus,
Coma, and the presently studied Hydra I, are not identical; in particular,
the slope seems to be steeper in poorer clusters.
\item Although the slope of the LF shows no larger differences between the
central and peripheral region, a small difference exist in 
the number of faint galaxies ($M\gtrsim-14$) in the two regions. 
In this range, the number of dwarfs in the cluster central region increases by 
about $40$\% compared to that in the peripheral region.
\item The LFs of red and blue galaxies, which are separated in the color-magnitude diagram,
show that (1) the Hydra I cluster is dominated by red galaxies in
both central and peripheral regions down to $-10$ mag,
(2) the variation of the shape of the LF in the faint magnitude range is due to
red rather than blue dwarfs, and
(3) blue dwarf galaxies do not strongly depend on the environment compared to red dwarf galaxies. 
\item The upturn in the total and red LFs at $M\sim-16$ is unique and common
to cluster environments, since field LFs show no such upturns.
The upturn may be caused by strong tidal interactions (such as harassment) in clusters.
\end{itemize}

%%%%%%%%%%%%%%%%%%%%%%%%%%%%%%%%%%%%%%%%%%%%%%%%%%%%%%
\acknowledgments

We are grateful to the staff of the Subaru Telescope for their assistance.
Data analysis was carried out on the general common use computer system at the Astronomical
Data Analysis Center, ADAC, of the National Astronomical Observatory of Japan.
M.T. acknowledges support from the Japan Society for the Promotion of
Science (JSPS) through JSPS research fellowships for Young Scientists.
Funding for the SDSS and SDSS-II was provided by the Alfred P. Sloan Foundation, 
the Participating Institutions, the National Science Foundation, the U.S. Department of Energy, 
the National Aeronautics and Space Administration, the Japanese Monbukagakusho, 
the Max Planck Society, and the Higher Education Funding Council for England. 
The SDSS Web site is http://www.sdss.org/.
This research has made use of the NASA/IPAC Extragalactic Database (NED) 
which is operated by the Jet Propulsion Laboratory, California Institute of 
Technology, under contract with the National Aeronautics and Space Administration.
This research has made use of the X-Rays Clusters Database (BAX)
which is operated by the Laboratoire d'Astrophysique de Tarbes-Toulouse (LATT),
under contract with the Centre National d'Etudes Spatiales (CNES). 
This work has used observations from the Digitized Sky Survey (DSS).
The DSS was produced at the Space Telescope Science Institute 
under U.S. government grant NAG W-2166. 
The images of these surveys are based on photographic data obtained using 
the Oschin Schmidt Telescope on Palomar Mountain and the UK Schmidt Telescope. 
The plates were processed into the present compressed digital form with the permission 
of these institutions.

%
%
%
%
%

%%%%%%%%%%%%%%%%%%%%%%%%%%%%%%%%%%%%%%%%%%%%%
%%%%%%%%%%%%%%%%%%%%%%%%%%%%%%%%%%%%%%%%%%%%%
%%%%%%%%%%%%%%%%%%%%%%%%%%%%%%%%%%%%%%%%%%%%%
%%%%%%%%%%%%%%%%%%%%%%%%%%%%%%%%%%%%%%%%%%%%%
%%%%%%%%%%%%%%%%%%%%%%%%%%%%%%%%%%%%%%%%%%%%%

%%%%table 1
\clearpage
\begin{deluxetable}{ccccccc}
\tabletypesize{\scriptsize}
\tablecaption{Observational Data\label{tbl-1}}
\tablewidth{0pt}
\tablehead{
\colhead{Region} & \colhead{R.A.} & \colhead{Dec.} & \colhead{Band\tablenotemark{*}} &
\colhead{Exp. Times} & \colhead{Lim. Mag.} & \colhead{Seeing}\\
\colhead{} & \colhead{(J2000.0)} & \colhead{(J2000.0)} & \colhead{} &
\colhead{(sec$\times$shots)} & \colhead{(5$\sigma$)} & \colhead{(arcsec)}
}
\startdata
Central  & $10^\mathrm{h}36^\mathrm{m}13^\mathrm{s}$  & $-27^\circ25'04\,''$  &  $B$   & $600\times3$  & 25.9 & 1.1\\
         &                  &                    & $R_C$  & $300\times4$  & 25.1 & 1.0\\
         &                  &                    &  $B$   & $120\times4$  & 24.7 & 1.1\\
         &                  &                    & $R_C$  & $~60\times4$   & 24.3 & 1.1\\
Peripheral & $10^\mathrm{h}35^\mathrm{m}22^\mathrm{s}$  & $-28^\circ24'30\,''$  &  $B$   & $600\times3$  & 26.0 & 1.1\\
         &                  &                    & $R_C$  & $300\times4$  & 25.5 & 1.0\\
         &                  &                    &  $B$   & $120\times4$  & 23.9 & 1.2\\
         &                  &                    & $R_C$  & $~60\times4$   & 24.5 & 1.1\\         
\enddata
\tablenotetext{*}{The central wavelengths are 440nm for the $B$-band and 650nm for the $R_C$-band.
The $R_C$-band is denoted as $R$-band for simplicity in this paper. 
}
\end{deluxetable}

%
%
%

%%%%table 2
\clearpage
\begin{deluxetable}{ccccccc}
\tabletypesize{\scriptsize}
\tablecaption{Faint-end LF of Nearby Clusters\label{tbl-2}}
\tablewidth{0pt}
\tablehead{
\colhead{Cluster} & \colhead{z} & \colhead{$L_X$\tablenotemark{*}} & \colhead{$\alpha$} &
\colhead{Mag. Range} & \colhead{Area} & \colhead{Reference}\\
\colhead{} & \colhead{} & \colhead{($10^{44}$ erg s$^{-1}$)} & \colhead{} &
\colhead{} & \colhead{(Mpc$^2$)} & \colhead{}
}
\startdata
Virgo  & $0.0036$\tablenotemark{\ a} & $0.21$\tablenotemark{\ d} &  $-1.6\pm0.1$  & $-15\lesssim M_B\lesssim -11$ & $1.69$ & \citealt{Sab03}\\
       &                             &                           & $-2.18\pm0.12$ & $-15\lesssim M_R\lesssim -11$ & $0.09$ & \citealt{Phi98}\\
Fornax & $0.0046$\tablenotemark{\ b} & $0.04$\tablenotemark{\ e} &    $\sim-2$    &            $M_B\lesssim -12$  & $1.69$ & \citealt{Kamb00}\\
Hydra  & $0.0126$\tablenotemark{\ c} & $0.28$\tablenotemark{\ e} & $-1.59\pm0.03$ &        $-19<M_B<-10$          & $0.09$ & This work\\
       &                             &                           & $-1.64\pm0.02$ &        $-21<M_R<-10$          & $0.09$ & This work\\
       & $0.0126$\tablenotemark{\ c} & $0.28$\tablenotemark{\ e} & $-1.56\pm0.04$ &        $-16<M_B<-10$          & $0.09$ & This work\\
       &                             &                           & $-1.60\pm0.03$ &        $-16<M_R<-10$          & $0.09$ & This work\\
Perseus& $0.0179$\tablenotemark{\ c} & $8.31$\tablenotemark{\ e} & $-1.44\pm0.04$ & $-20\lesssim M_B\lesssim -11$ & $0.09$ & \citealt{Con02}\\
Coma   & $0.0231$\tablenotemark{\ c} & $4.04$\tablenotemark{\ e} & $-1.42\pm0.05$ & $-19\lesssim M_R\lesssim -11$ & $0.04$ & \citealt{Ber95}\\
       &                             &                           & $-1.41\pm0.05$ & $-19\lesssim M_R\lesssim -12$ & $0.49$ & \citealt{Sec97}\\         
\enddata
\tablenotetext{*}{$L_X$ of each clusters in the 0.1-2.4 keV band
}
\tablenotetext{a}{\citealt{Ebe98}
}
\tablenotetext{b}{\citealt{Abel89}
}
\tablenotetext{c}{\citealt{SR99}
}
\tablenotetext{d}{\citealt{Matsu00}
}
\tablenotetext{e}{\citealt{RB02}
}
\end{deluxetable}

%
%
%

%%%%figure E-1
\clearpage
%%%%

%%%%figure E-2
\clearpage
%%%%

%%%%figure E-3
\clearpage
%%%%

%%%%figure E-4
\clearpage
%%%%

%%%%figure E-5
\clearpage
%%%%

%%%%figure E-6
\clearpage
%%%%

%%%%%%%%%%%%%%%%%%%%%%%%%%%%%%%%
%%%%%%%%%%%%%%%%%%%%%%%%%%%%%%%%

\begin{thebibliography}{}

\bibitem[Abell et al.(1989)]{Abel89} Abell, G.~O., Corwin, 
H.~G., Jr., \& Olowin, R.~P.\ 1989, \apjs, 70, 1 

\bibitem[Adami et al.(2006)]{Adam06} Adami, C., et al.\ 2006, 
\aap, 459, 679 

\bibitem[Adelman-McCarthy et al.(2006)]{AM06} 
Adelman-McCarthy, J.~K., et al.\ 2006, \apjs, 162, 38

\bibitem[Andreon \& Cuillandre(2002)]{AC02} Andreon, S., \& 
Cuillandre, J.-C.\ 2002, \apj, 569, 144

\bibitem[Andreon et al.(2006)]{And06} Andreon, S., 
Cuillandre, J.-C., Puddu, E., \& Mellier, Y.\ 2006, \mnras, 372, 60 

\bibitem[Arnouts et al.(2001)]{Arn01} Arnouts, S., et al.\ 
2001, \aap, 379, 740 
 
\bibitem[Baier \& Oleak(1983)]{BO83} Baier, F.~W., \& Oleak, 
H.\ 1983, Astronomische Nachrichten, 304, 277 

\bibitem[Bekki et al.(2001)]{Bekk01} Bekki, K., Couch, W.~J., 
\& Drinkwater, M.~J.\ 2001, \apjl, 552, L105 

\bibitem[Bernstein et al.(1995)]{Ber95} Bernstein, G.~M., 
Nichol, R.~C., Tyson, J.~A., Ulmer, M.~P., \& Wittman, D.\ 1995, \aj, 110, 
1507 

\bibitem[Bertin \& Arnouts(1996)]{BA96} Bertin, E., \& 
Arnouts, S.\ 1996, \aaps, 117, 393 

\bibitem[Bertin \& Dennefeld(1997)]{BD97} Bertin, E., \& 
Dennefeld, M.\ 1997, \aap, 317, 43 

\bibitem[Blanton et al.(2005)]{Bla05} Blanton, M.~R., Lupton, 
R.~H., Schlegel, D.~J., Strauss, M.~A., Brinkmann, J., Fukugita, M., \& 
Loveday, J.\ 2005, \apj, 631, 208

\bibitem[Capak et al.(2004)]{Capa04} Capak, P., et al.\ 2004, 
\aj, 127, 180 

\bibitem[Christlein \& Zabludoff(2003)]{CZ03} Christlein, 
D., \& Zabludoff, A.~I.\ 2003, \apj, 591, 764 

\bibitem[Conselice et al.(2002)]{Con02} Conselice, C.~J., 
Gallagher, J.~S., \& Wyse, R.~F.~G.\ 2002, \aj, 123, 2246 

\bibitem[De Propris et al.(2003)]{DeP03} De Propris, R., et 
al.\ 2003, \mnras, 342, 725 

\bibitem[Drinkwater et al.(1999)]{Drink99} Drinkwater, M.~J., 
Phillipps, S., Gregg, M.~D., Parker, Q.~A., Smith, R.~M., Davies, J.~I., 
Jones, J.~B., \& Sadler, E.~M.\ 1999, \apjl, 511, L97 

\bibitem[Drinkwater et al.(2003)]{Drink03} Drinkwater, M.~J., 
Gregg, M.~D., Hilker, M., Bekki, K., Couch, W.~J., Ferguson, H.~C., Jones, 
J.~B., \& Phillipps, S.\ 2003, \nat, 423, 519 

\bibitem[Ebeling et al.(1998)]{Ebe98} Ebeling, H., Edge, 
A.~C., Bohringer, H., Allen, S.~W., Crawford, C.~S., Fabian, A.~C., Voges, 
W., \& Huchra, J.~P.\ 1998, \mnras, 301, 881 

\bibitem[Ferguson \& Sandage(1988)]{FS88} Ferguson, H.~C., 
\& Sandage, A.\ 1988, \aj, 96, 1520 

\bibitem[Fitchett \& Merritt(1988)]{FM88} Fitchett, M., \& 
Merritt, D.\ 1988, \apj, 335, 18 

\bibitem[Fukugita et al.(1996)]{Fuku96} Fukugita, M., 
Ichikawa, T., Gunn, J.~E., Doi, M., Shimasaku, K., \& Schneider, D.~P.\ 
1996, \aj, 111, 1748 

\bibitem[Heydon-Dumbleton et al.(1989)]{HD89} 
Heydon-Dumbleton, N.~H., Collins, C.~A., \& MacGillivray, H.~T.\ 1989, 
\mnras, 238, 379 

\bibitem[Hilker et al.(2003)]{Hil03} Hilker, M., Mieske, S., 
\& Infante, L.\ 2003, \aap, 397, L9 

\bibitem[Iye et al.(2004)]{Iye04} Iye, M., et al.\ 2004, 
\pasj, 56, 381 

\bibitem[Jones et al.(2006)]{Jone06} Jones, D.~H., Peterson, 
B.~A., Colless, M., \& Saunders, W.\ 2006, \mnras, 369, 25 

\bibitem[Kambas et al.(2000)]{Kamb00} Kambas, A., Davies, 
J.~I., Smith, R.~M., Bianchi, S., \& Haynes, J.~A.\ 2000, \aj, 120, 1316 

\bibitem[K{\"u}mmel \& Wagner(2001)]{KW01} K{\"u}mmel, 
M.~W., \& Wagner, S.~J.\ 2001, \aap, 370, 384 

\bibitem[Landolt(1992)]{Lan92} Landolt, A.~U.\ 1992, \aj, 
104, 340

\bibitem[Lee \& Shandarin(1999)]{LS99} Lee, J., \& 
Shandarin, S.~F.\ 1999, \apjl, 517, L5 

\bibitem[Matsumoto et al.(2000)]{Matsu00} Matsumoto, H., Tsuru, 
T.~G., Fukazawa, Y., Hattori, M., \& Davis, D.~S.\ 2000, \pasj, 52, 153 

\bibitem[Miller \& Owen(2003)]{MO03} Miller, N.~A., \& Owen, 
F.~N.\ 2003, \aj, 125, 2427 
 
\bibitem[Miyazaki et al.(2002)]{Miya02} Miyazaki, S., et al.\ 
2002, \pasj, 54, 833

\bibitem[Moore et al.(1996)]{Moo96} Moore, B., Katz, N., 
Lake, G., Dressler, A., \& Oemler, A., Jr.\ 1996, \nat, 379, 613

\bibitem[Moore et al.(1998)]{Moo98} Moore, B., Governato, F., 
Quinn, T., Stadel, J., \& Lake, G.\ 1998, \apjl, 499, L5 

\bibitem[Ouchi et al.(2004)]{Ouc04} Ouchi, M., et al.\ 2004, 
\apj, 611, 660 

\bibitem[Owen et al.(2005)]{Owen05} Owen, F.~N., Ledlow, 
M.~J., Keel, W.~C., Wang, Q.~D., \& Morrison, G.~E.\ 2005, \aj, 129, 31 

\bibitem[Phillipps et al.(1998)]{Phi98} Phillipps, S., 
Parker, Q.~A., Schwartzenberg, J.~M., \& Jones, J.~B.\ 1998, \apjl, 493, 
L59 

\bibitem[Popesso et al.(2005)]{Pop05} Popesso, P., 
B{\"o}hringer, H., Romaniello, M., \& Voges, W.\ 2005, \aap, 433, 415 

\bibitem[Press \& Schechter(1974)]{PS74} Press, W.~H., \& 
Schechter, P.\ 1974, \apj, 187, 425 

\bibitem[Reiprich \& B{\"ouml}hringer(2002)]{RB02} Reiprich, 
T.~H., \& B{\"ouml}hringer, H.\ 2002, \apj, 567, 716 

\bibitem[Richter et al.(1982)]{Ric82} Richter, O.-G., 
Huchtmeier, W.~K., \& Materne, J.\ 1982, \aap, 111, 193 

\bibitem[Richter(1987)]{Ric87} Richter, O.-G.\ 1987, \aaps, 
67, 261 

\bibitem[Richter(1989)]{Ric89} Richter, O.-G.\ 1989, \aaps, 
77, 237 

\bibitem[Sabatini et al.(2003)]{Sab03} Sabatini, S., Davies, 
J., Scaramella, R., Smith, R., Baes, M., Linder, S.~M., Roberts, S., \& 
Testa, V.\ 2003, \mnras, 341, 981 

\bibitem[Schechter(1976)]{Sch76} Schechter, P.\ 1976, \apj, 
203, 297

\bibitem[Schlegel et al.(1998)]{Sgel98} Schlegel, D.~J., 
Finkbeiner, D.~P., \& Davis, M.\ 1998, \apj, 500, 525  

\bibitem[Secker et al.(1997)]{Sec97} Secker J., Harris W. E. \& Plummer J. D. 
1997, PASP, 109, 1377

\bibitem[Sekiguchi et al. in prep.()]{Sek06} Sekiguchi, K., \& 
SXDS, in prepare  

\bibitem[Struble \& Rood(1991)]{SR91} Struble, M.~F., \& 
Rood, H.~J.\ 1991, \apjs, 77, 363  

\bibitem[Struble \& Rood(1999)]{SR99} Struble, M.~F., \& 
Rood, H.~J.\ 1999, \apjs, 125, 35 

\bibitem[Trentham et al.(2005)]{Tre05} Trentham, N., Sampson, 
L., \& Banerji, M.\ 2005, \mnras, 357, 783 

\bibitem[White \& Rees(1978)]{WR78} White, S.~D.~M., \& 
Rees, M.~J.\ 1978, \mnras, 183, 341 

\bibitem[White \& Frenk(1991)]{WF91} White, S.~D.~M., \& 
Frenk, C.~S.\ 1991, \apj, 379, 52 

\bibitem[Yagi et al.(2002a)]{YK02a} Yagi, M., Kashikawa, N., 
Sekiguchi, M., Doi, M., Yasuda, N., Shimasaku, K., \& Okamura, S.\ 2002, 
\aj, 123, 66 
 
\bibitem[Yagi et al.(2002b)]{YK02b} Yagi, M., Kashikawa, N., 
Sekiguchi, M., Doi, M., Yasuda, N., Shimasaku, K., \& Okamura, S.\ 2002, 
\aj, 123, 87 

\bibitem[Yasuda et al.(2001)]{Yasu01} Yasuda, N., et al.\ 
2001, \aj, 122, 1104 

\bibitem[York et al.(2000)]{Yor00} York, D.~G., et al.\ 2000, 
\aj, 120, 1579 


\end{thebibliography}
\end{document}